\documentclass[%
reprint,
amsmath,amssymb,
aps,
]{revtex4-1}

\usepackage{graphicx}
\usepackage{dcolumn}
\usepackage{bm}
\usepackage{subfigure}
\usepackage{mathtools}
\usepackage{bbm}
\usepackage{amsfonts}
\usepackage{amssymb}

\def\bh{{\mathbb{H}}}

\def\bb{{\mathbb{B}}}

\def\bf{{\mathbb{F}}}

\def\pd{{p_{\mathbf{d}}}}

\def\bd{{{\mathbf{d}}}}

\def\bm{{{\mathbf{m}}}}
\def\bbE{{{\mathbb{E}}}}

\def\mn{{\mathcal{N}}}

\newcommand{\lsb}{\left(}
\newcommand{\rsb}{\right)}
\newcommand{\lmb}{\left[}
\newcommand{\rmb}{\right]}
\newcommand{\lab}{\left<}
\newcommand{\rab}{\right>}

\newcommand{\qrol}{q_{r, 1, \ell}}
\newcommand{\qrtl}{q_{r, 2, \ell}}
\newcommand{\qbol}{q_{b, 1, \ell}}
\newcommand{\qbtl}{q_{b, 2, \ell}}

\newcommand{\qrolOne}{q_{r, 1, \ell + 1}}
\newcommand{\qrtlOne}{q_{r, 2, \ell + 1}}
\newcommand{\qbolOne}{q_{b, 1, \ell + 1}}
\newcommand{\qbtlOne}{q_{b, 2, \ell + 1}}

\newcommand{\qrolInf}{q_{r, 1, \infty}}
\newcommand{\qrtlInf}{q_{r, 2, \infty}}
\newcommand{\qbolInf}{q_{b, 1, \infty}}
\newcommand{\qbtlInf}{q_{b, 2, \infty}}

\newcommand{\dr}{d_r}
\newcommand{\db}{d_b}

\newcommand{\grOne}{g_{r, 1}(x)}
\newcommand{\grTwo}{g_{r, 2}(x)}
\newcommand{\gbOne}{g_{b, 1}(x)}
\newcommand{\gbTwo}{g_{b, 2}(x)}

\newcommand{\tina}{\textit{inactive}}
\newcommand{\ta}{\textit{active}}

\newcommand{\tr}{\textit{red}}
\newcommand{\tb}{\textit{blue}}
\newcommand{\tg}{\textit{global}}

\begin{document}

%

\title{Multi-Stage Complex Contagions in Random Multiplex Networks}

\author{Yong Zhuang and Osman Ya\u{g}an}
\affiliation{Department of ECE, Carnegie Mellon University, Pittsburgh, PA 15213, USA}
\date{\today}


\begin{abstract}

Complex contagion models have been developed to understand a wide range of social phenomena such as
adoption of cultural fads, the diffusion of belief, norms, and innovations in social networks, and the rise of collective action to join a riot.
Most existing works focus on contagions where individuals' states  are represented by {\em binary} variables, and propagation takes place over a single isolated network.
However, characterization of an individual's standing on a given matter as a binary state might be overly simplistic as most of our opinions, feelings, and perceptions vary over more than two states.
Also, most real-world contagions take place over multiple networks (e.g., Twitter and Facebook) or involve {\em multiplex} networks where individuals engage in different {\em types} of relationships (e.g., acquaintance, co-worker, family, etc.).
To this end, this paper studies {\em multi-stage} complex contagions that take place over multi-layer or multiplex networks.
Under a linear threshold based contagion model, we give analytic results for the probability and expected size of $\tg$ cascades, i.e., cases where a randomly chosen node can initiate a propagation that eventually reaches a {\em positive} fraction of the whole population.
Analytic results are also confirmed and supported by an extensive numerical study.
In particular, we demonstrate how the dynamics of complex contagions is affected by the extra weight exerted by \textit{hyper-active} nodes and by the structural properties of the networks involved. Among other things, we reveal an interesting connection between the assortativity of a network and the impact of \textit{hyper-active} nodes on the cascade size.

\end{abstract}

\maketitle


\section{INTRODUCTION}

Modeling and analysis of dynamical processes in complex networks has been a very active research field in the past decade. This has led to many advances in our understanding and ability to control a wide range of physical and social phenomena. Examples
include adoption of cultural fads, the diffusion of beliefs, norms, and innovations in social networks 
\cite{watts2002simple, dodds2009analysis,yagan2013conjoining, yaugan2012analysis,DuanChenLiuJin,Dajun_RealTime, zhuang2016information}, 
disease contagion in human and animal populations \cite{ AndersonMay,newman2002spread,Basar1,Basar2,pare2015stability}, 
cascading failures in {\em interdependent} infrastructures 
\cite{buldyrev2010catastrophic,Vespignani,YaganQianZhangCochranLong},
{\em insolvency} and {\em default} cascades in financial networks \cite{hurd2011framework,hurd2016double},
and the  spread of computer viruses 
or worms on the Web \cite{NewmanForrestBalthrop,BalthropForrestNewmanWilliamson}.

In this work, we focus on {\em complex} contagions, a class of dynamical processes typically used in modeling the propagation of {\em influence} in social networks. In particular, complex contagion models are used when {\em social reinforcement} plays an important role in the propagation process, i.e., when {\em multiple} sources of exposure is needed for an individual to adopt an activity. Examples include the spread of social movements and radical behavior, the rise of collective action to join a riot, or the decision to support one political candidate versus the other. 
This differs from the class of models known as {\em simple} contagions, where propagation often takes place after only a single copy is received; e.g., spread of diseases, viruses, etc.

Complex contagions have typically been studied in the literature using a linear threshold model.
The original threshold model was proposed by Watts \cite{watts2002simple} for binary-state dynamics.
In particular, each node is assumed to be in one of two states, {\em inactive} or {\em active}, and is initially given a threshold $\tau$ in $(0, 1]$.
Then, a randomly chosen node is set as {\em active}, while all others are {\em inactive}.
At any point in time, if an {\em inactive} node has $d$ neighbors of which $m$ are {\em active}, we determine if it will be activated by checking the relationship between $\frac{m}{d}$ and the pre-assigned threshold $\tau$.
If $\frac{m}{d} \geq \tau$, then the node will turn $\ta$.
Otherwise, if $\frac{m}{d} < \tau$, it stays $\tina$. It is also assumed that once activated, a node will remain active forever.

The Watts threshold model focuses solely on \textit{single-stage} complex contagions, i.e., it is suitable only for binary-state dynamics.
However, the characterization of an individual's standing on a given matter as a binary state might be overly simplistic, since most of our opinions, feelings, and perceptions vary over more than two states.
Similarly, followers of a radical organization or a revolutionary movement may have varying levels of commitment to the cause, or might have varying desire and ability to recruit new members.
Thus, it is of interest to study a model where nodes exhibit a richer set of states.
To this end, Melnik et al. \cite{melnik2013multi} 
generalized the linear threshold model and introduced a \textit{multi-stage} contagion model.   There, nodes can be inactive or can be in one of several levels of active states (e.g., active, hyper-active, etc.). 
The details of this model are given in Section \ref{sec:multi_stage_models}.

In this work, we aim to extend the literature on multi-stage contagions from single networks to multi-layer and multiplex networks.
Our intuition is that most real-world influence propagation events take place over multiple networks.
For example, individuals may participate in multiple online social networks (e.g., Facebook, Twitter, etc.), and may have different levels of influence in each network.
Similarly, within a single network, individuals may form different types of relationships (e.g., friendship, colleagueship, kinship, etc.), and each relationship type might have a different impact on propagation of influence in a given context. For example, video games might be more likely to spread among high-school friends rather than parents,
while the opposite might be true for political ideas. 
Thus, if we do not distinguish different types of relationships, dynamics of influence propagation may not be accurately captured.
To address this issue, Ya\u{g}an and Gligor proposed  \cite{yaugan2012analysis} a {\em context-dependent} linear threshold model in {\em multiplex} networks.
In this model, each link type has its own weight on the propagation of the influence (which might change from one context to another); 
the detailed explanation of the model can be found in Section \ref{sec:multi_stage_models}.
However, the model in \cite{yaugan2012analysis}  considers only binary-state dynamics and hence is not able to model multi-stage contagions (where nodes can belong to a rich set of states).

The discussion above indicates that understanding how influence propagates in real-world networks requires incorporating two important factors: i) the multi-state nature of individuals' activity levels; and ii) the multilayer/multiplex nature of the networks involved in the propagation.
To this end, in this work, we develop and analyze a multi-stage contagion model on multiplex networks.
For simplicity, we assume that there are two types of links, $\tr$ and $\tb$ in the network, and three stages, {\em inactive}, {\em active}, and {\em hyper-active}, representing the possible node states in the contagion dynamics. 
Our main results are i) calculating the probability of triggering {\em global cascades}, 
i.e., cases where a positive fraction of nodes (in the asymptotic limit) eventually becomes active or hyper-active when a randomly selected node is switched to the active state; ii) calculating the expected size of global cascades when they are possible. 
This is done for networks generated {\em randomly} from a given degree distribution, i.e., networks generated by the configuration model \cite{newman2001random}.

Our analytic  results  are   confirmed  and supported by an extensive numerical study. In particular, we demonstrate how the extra weight of hyper-active nodes in the multi-stage model and the structural properties of networks affect the contagion dynamics.
For instance, a particularly interesting scenario is when the hyper-active state is manifested in only one link type. This is motivated by the case where people may be more willing to express their viewpoints to close friends instead of office-mates, or may be more influential in one social network (e.g., Twitter) versus another (e.g., Facebook).
Among other things, we reveal interesting connections between the assortativity of a network and the impact of \textit{hyper-active} nodes on cascade size.
For instance, when the network is highly assortative, the influence exerted by the hyper-active nodes may change not only the critical transition points, but  also the number and {\em order} of transitions; while the affect is much more limited in networks with low assortativity.


The rest of the paper is organized as follows.
In Section \ref{sec:models}, we introduce the network and contagion models.
Then, we describe the problem of interest and our main results in Section \ref{sec:results}.
In Section \ref{sec:numerical_results}, we present numerical results that demonstrate the accuracy of our analysis in the finite node regime, and discuss the impact of hyper-influencers on complex contagions.
We conclude the paper in Section \ref{sec:conclusion} where we also suggest several directions for future work.

\section{Model Definition:Networks and Dynamics}
\label{sec:models}
In this section, we first present the multiplex network model considered in this work, and then describe the multi-stage complex contagion model.
\subsection{Multi-layer and multiplex network models}
\label{sec:multilayer_multiplex_network_models}
We consider multiplex networks where links are classified into different types (or, colors).
For convenience, in the following discussion, we focus on a multiplex network with two types of links, \textit{red} and \textit{blue}, but the model and results can be easily extended to an arbitrary number of link types.
These two link types can be motivated by the case where one color accounts for edges in Facebook while the other for edges in Instagram.
Alternatively, one link color may be representing close friendship links while the other representing ``acquaintances" in a social network.
In this network model, we let $\mn = \{1, 2, \dots, n\}$ denote the vertex set, with $n$ standing for the number of nodes.
We let $\mn_r \subset \mn$ denote the set of vertices that have \textit{red} edges and $\mn_b \subset \mn$ denote the set of vertices having \textit{blue} edges.
For simplicity, we assume $\mn_b = \mn$, which means all vertices in the network may have blue edges.
To model the possibility that not everyone may have red links, we assume that each vertex in $\mn$ has red links with probability $\alpha \in (0, 1]$:
\begin{align}
    \mathbb{P}[i \in \mn_r] = \alpha, \quad i = 1, \dots, n.
\end{align}

\begin{figure*}[!t]
    \centering
    \subfigure[]{
        \hspace{0cm}
        \includegraphics[width=0.45\textwidth]{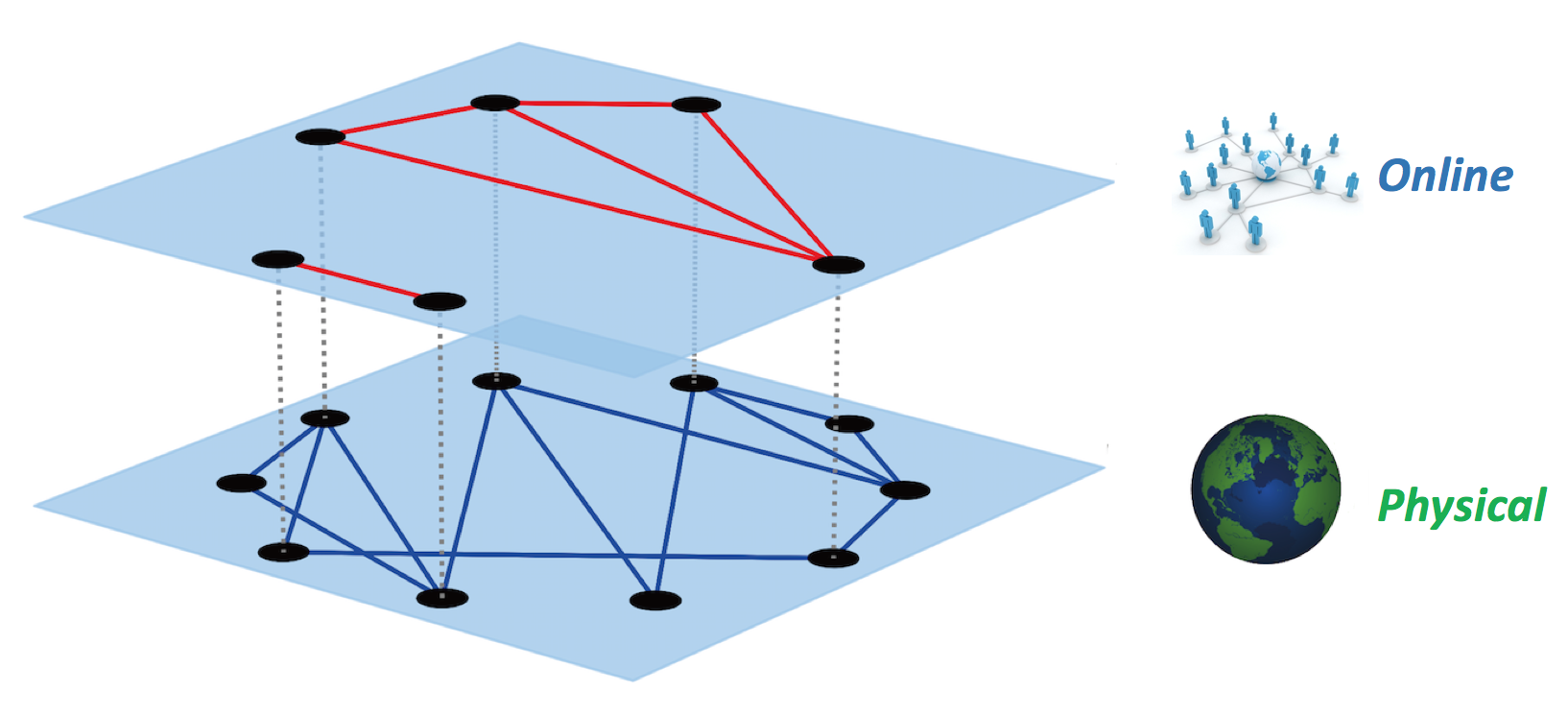}
        \label{fig:multilayer_networks}
    }
    \hspace{2em}
    \subfigure[]{
        \hspace{0cm} 
        \includegraphics[width=0.45\textwidth] {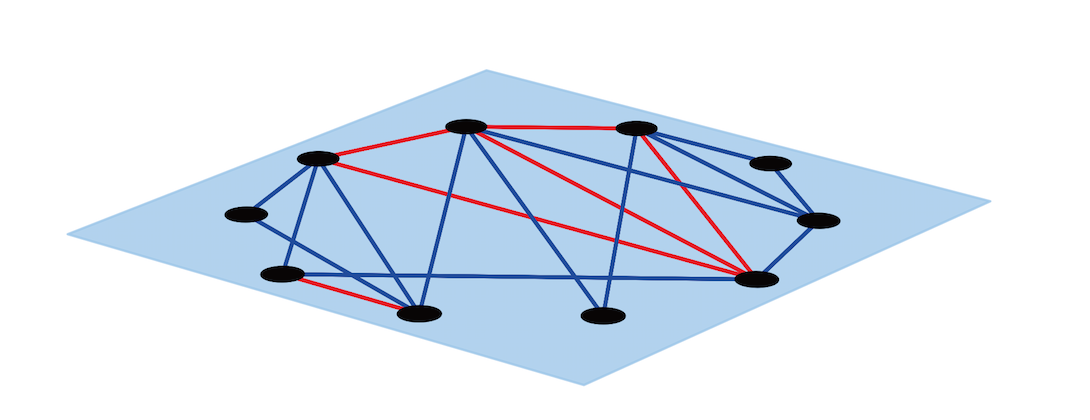} 
        \label{fig:multiplex_network}
    }
    \caption{\sl Illustration of a multi-layer and a multiplex network representation of our model.
    In (a), we see a multi-layer network (e.g., a Physical communication layer and an online social network layer) with overlapping vertex sets; vertical dashed lines represent nodes corresponding to the same individual.
    In (b), we see the equivalent representation of this model by a multiplex network.
    Edges from Facebook are shown in red and edges from the physical network are shown in blue.
    }
    \label{fig:network_models}
\end{figure*}

This network model can be interpreted in two different ways.
The first one is a multi-layer network where each network layer is generated by the widely used configuration model \cite{newman2001random,molloy1995critical,bollobas2001cambridge}; this  case is illustrated in Figure \ref{fig:multilayer_networks}.
In particular, we use $P(d_r)$ (resp. $P(d_b)$) as the {\em degree distribution} to determine the number of red (resp. blue) edges that will be assigned to each node in $\mn_r$ (resp. $\mn_b$). Once the degree of each node is determined, we generate the networks $\mathbb{R}$ and $\mathbb{B}$ by selecting a graph uniformly at random from among all possible graphs that have the same degree sequence; see  \cite{newman2001random,molloy1995critical} for more details. 
Next, we take a union of the edges in $\mathbb{R}$ and $\bb$ to create a network $\bh$.
Equivalently, we can consider a multiplex network model generated by the {\em colored} configuration model \cite{soderberg2003random}.
Let $\mathbf{d}=(d_r, d_b)$ 
denote the colored degree of a node, where $d_r$ and $d_b$ stand for the number of red edges and blue edges incident on it.
Each of the $n$ nodes in the network is assigned a colored degree by independently drawing from the distribution $P_{\bd}$. 
Then, pairs of edges of the same color are randomly chosen and connected together until none is left; see \cite{soderberg2003random} for details.
Figure \ref{fig:multiplex_network} is an illustration of this multiplex network model.

\vspace{-2mm}

\subsection{Multi-stage content-dependent linear threshold model for complex contagions}
\label{sec:multi_stage_models}
We first introduce the single-stage content-dependent linear threshold model \cite{yaugan2012analysis} which is a generalization of the vanilla threshold model \cite{watts2002simple}.
In the content-dependent linear threshold model, links are classified into $r$ types.
For a given content (a view, rumor, product, etc.), 
scalars $c_i$, $i=1,\ldots, r$ 
represent the weight (i.e., relative importance) of type-$i$ edges on spreading this particular content.
Nodes belong to either one of the two states, active or inactive, and each node is assigned a threshold $\tau$ in $(0, 1]$ drawn from a distribution $P(\tau)$.
Given an inactive node with $m_{i}$ active and $d_i - m_i$ inactive neighbors for each link type-$i$, $i = 1, \dots, r$, an inactive node will turn active if $\frac{\sum_{i} c_i m_i}{\sum_{i} c_i d_i} \geq \tau$.
Namely, an inactive node with $\mathbf{m} = (m_1, \dots, m_r)$ and $\mathbf{d} = (d_1, \dots, d_r)$ will turn $\ta$ with probability
\begin{align}
    F[\bm, \bd] \triangleq \mathbb{P}\left[\frac{\sum_{i=1}^{r} c_i m_i}{\sum_{i=1}^{r} c_i d_i} \ge \tau\right].
\end{align}
Throughout, $F[\bm, \bd]$ is referred to as the response function.  
If we do not distinguish the edge types or simply set $c_i = 1$ for all $i = 1, \dots, r$, then this model reduces to the Watts' threshold model \cite{watts2002simple}.
This content-dependent threshold model enables us to model the case where people's influence on others vary according to their relationship type, or the social network that they are interacting through.

Different from the single-stage threshold model where nodes can only be in two states, the multi-stage linear threshold model \cite{melnik2013multi} allows nodes to be in a richer set of active states.
In this work, we assume that nodes can belong to three states, inactive, active, and hyper-active.
In the following discussion, we use state-0, state-1, and state-2 to represent the inactive, active, and hyper-active state, respectively.
Let $\tau_1$ and $\tau_2$ denote the thresholds associated with transitioning to the active and hyper-active states, respectively.
The hyper-active individuals are assumed to be $\beta$-times more influential than active nodes in the propagation process (where $\beta \geq 1$).
For example, an individual with $d$ neighbors of which $m_{1}$ are active and $m_{2}$ are hyper-active, the probability of switching to state-$i$ from the inactive state (i.e., state-0) is given by:
\begin{align}
F_i[\bm, d] \triangleq \mathbb{P}\left[\tau_{i} \leq  \frac{m_1 + \beta m_2}{d} \leq \tau_{i+1}\right], \quad i = 0, 1, 2, 
    \label{eq:response_function_multistage1}
\end{align}
where $\bm = (m_{1}, m_{2})$, $\tau_0 = 0$, $\tau_3 = \infty$, and $\beta \geq 1$.
Although we assume there are three states in the contagion process, our analysis can be extended to an arbitrary number of states.

Finally, we introduce the multi-stage content-dependent linear threshold model.
Assume that there are two types of links, red and blue, in the network, and that nodes can be in three states, inactive, active, and hyper-active.
We let $c_r$ and $c_b$ denote the weight of red and blue edges, respectively, and set $c = \frac{c_r}{c_b}$.
With this notation, the probability of an inactive node switching to state-$i$ is given by:
\begin{align}
&F_i[\bm, \bd]  \label{eq:response_function_multistage2} \\
& \triangleq \mathbb{P}\left[\tau_i \leq \frac{c(m_{r, 1} + \beta m_{r, 2}) + m_{b, 1} + \beta m_{b, 2}}{c d_{r} + d_{b}} \leq \tau_{i+1}\right],
    \nonumber
\end{align}
where $\bm = (m_{r, 1}, m_{r, 2}, m_{b, 1}, m_{b, 2})$, $\bd = (d_r, d_b)$, $m_{r, 1}$ and $m_{r, 2}$ (resp. $m_{b, 1}$ and $m_{b, 2}$) denote the number of active and hyper-active neighbors connected through a red (resp. blue) edge, and $d_r$ and $d_b$ denote the number of red and blue neighbors respectively.

\section{Main Results}
\label{sec:results}
We assume that the contagion process starts by randomly choosing an initial node and setting it as \textit{active}, while all other nodes are in the inactive state. 
The influence might then propagate in the network according to (\ref{eq:response_function_multistage2}) and other nodes might turn active, and so on. Since the contagion process is monotone (i.e., an active node can never switch back to inactive), it will eventually stop, i.e., a steady-state will be reached.

Our main goals are i) determining the conditions (in terms of network parameters) for {\em global cascades} to be possible, i.e., cases where influence starts from a single individual (selected uniformly at random) and eventually reaches a positive fraction of the population in the limit of large network sizes; ii) calculating the expected size of global cascades when they are possible; and iii) calculating the probability of triggering global cascades.

\subsection{Expected cascade size and the condition to have a \textit{global} cascade}
We start the analysis with computing the expected size of global cascades when they occur.
Consider a random variable $S$ defined as
\begin{align*}
    S \triangleq \frac{\text{ \# of active and hyper-active nodes at steady-state}}{n},
\end{align*}
where $n$ is the number of nodes in the network. Then, a {\em global} cascade is said to take place if $S>0$ in the limit $n \to \infty$, and our main goal is to 
derive
\begin{align*}
   \lim_{n \to \infty} \mathbb{E}\left[S ~|~ S > 0 \right],
\end{align*}
which gives the expected size of global cascades when they  exist. 
Our analysis is based on the ``tree-approximation" approach \cite{gleeson2007seed, yaugan2012analysis, melnik2013multi}, which was developed to analyze the zero-temperature random-field Ising model on Bethe lattices \cite{sethna1993hysteresis}.
The tree-approximation approach assumes that the network has a locally tree-like structure.
Labeling the tree structure from the bottom to the top, it is assumed that the node states are updated starting from the bottom, and continuing to the top one level at a time.
In other words, the nodes at level $\ell$ will not update their states until the nodes at level $0, 1, \dots, \ell -1$ have finished updating. 
We define $\qrol$ (resp. $\qbol$) as the probability that the node at level $\ell$ is active and is connected to its only parent at level $\ell + 1$ by a red (resp. blue) edge.
Similarly, we define $\qrtl$ (resp. $\qbtl$) as the probability that an inactive node at level $\ell$, which is attached to its only parent via a red (resp. blue) edge, turns hyper-active.
We assume that the parent nodes at level $\ell + 1$ are inactive.

In the interest of brevity, we only introduce the derivation of $\qrolOne$, because the derivations of $\qrtlOne$, $\qbolOne$ and $\qbtlOne$ can be explained in a similar way.
Since $\qrolOne$ cannot be expressed explicitly, we derive a recursive relation in terms of $\qrol$, $\qrtl$, $\qbol$, and $\qbtl$; see (\ref{eq:qrolone})-(\ref{eq:qbtlone}). 
The validity of the expression  (\ref{eq:qrolone}) for 
$\qrolOne$ can be explained as follows.
Consider an inactive node at level $\ell + 1$ that is connected to its unique parent at level $\ell+2$ via a red edge, and that has colored degree $\bd = (d_r, d_b)$.
The probability that this node has $i$ active children connected via red edges, $s$ active children connected via blue edges, $j$ hyper-active children connected via red edges, and $t$ hyper-active children connected via blue edges, {\em and} that it turns active is given by
\begin{align}
    &{d_r - 1 \choose i} {d_r - 1 - i\choose j} \qrol^{i} \qrtl^{j} (1 - \qrol - \qrtl)^{d_r - 1 - i - j} \notag \\
    & \times {d_b \choose s} {d_b - s \choose t}  \qbol^{s} \qbtl^{t} (1 - \qbol - \qbtl)^{d_b - s - t} \notag \\
    & \times F_1\left[(i, j, s, t), \bd\right],
    \label{eq:qrolOne_single}
\end{align}
where $F_1\left[(i, j, s, t), \bd\right]$ is as defined in (\ref{eq:response_function_multistage2}); i.e., it denotes the probability that an inactive node with a colored degree $\bd$ and a group of active and hyper-active neighbors for each color represented by $\bm = (i, j, s, t)$ switches to state-$1$.
To simplify the notation, we use $\bf_1 \lmb (i, j, s, t), (x, y)\rmb$ as defined at (\ref{eq:f}), so the term given in (\ref{eq:qrolOne_single}) becomes equivalent to $\bf_1 \lmb (i, j, s, t), (\dr - 1, \db), \ell \rmb$.

The intuition behind (\ref{eq:qrolOne_single}) is as follows. Since we assume that the network is  tree-like, the state of each child node at level $\ell$ is independent from other children at the same level.
Thus, we multiply together the probability of being at a specific state for each child node to get the whole expression (\ref{eq:qrolOne_single}) using a simple combinatorial argument.
The reason behind using $d_r - 1$ rather than $d_r$ in (\ref{eq:qrolOne_single}) is the fact that the node under consideration is attached to its unique parent at level $\ell+2$ through a \textit{red} edge, and by assumption this parent node is inactive; recall that a node at level $\ell+2$ can not update its state until all nodes in level $\ell+1$ finish updating. 
A node that is known to have at least one red edge can be seen to have colored {\em degree} $\bd = (d_r, d_b)$
with probability $\frac{\dr\pd}{\lab\dr\rab}$; e.g., see \cite{newman2001random,yaugan2012analysis} for a discussion on the {\em excess} degree distribution.
Finally, we get the detailed expressions of $\qrolOne$ (\ref{eq:qrolone})
after
taking the expectation of (\ref{eq:qrolOne_single}) over the degree of the node at level $\ell + 1$.
We can use similar arguments to derive expressions for $\qrtlOne$, $\qbolOne$, and $\qbtlOne$.
The expressions of all four probabilities are shown in (\ref{eq:qrolone}) - (\ref{eq:qbtlone}).

\begin{widetext}
\begin{align}
    \qrolOne &= \sum_{\bd}\frac{\dr\pd}{\lab\dr\rab} \sum_{i = 0}^{\dr - 1} \sum_{j = 0}^{\dr - 1 - i} \sum_{s = 0}^{\db} \sum_{t = 0}^{\db - s} 
    \bf_1 \lmb (i, j, s, t), (d_r - 1, d_b), \ell \rmb
    \label{eq:qrolone}
    \\
    \qrtlOne &= \sum_{\bd} \frac{\dr\pd}{\lab \dr \rab} \sum_{i = 0}^{\dr - 1} \sum_{j = 0}^{\dr - 1 - i} \sum_{s = 0}^{\db} \sum_{t = 0}^{\db - s} \bf_2 \lmb (i, j, s, t), (d_r - 1, d_b), \ell \rmb
    \label{eq:qrtlone}
    \\
    \qbolOne &= \sum_{\bd}\frac{\db\pd}{\lab\db\rab} \sum_{i = 0}^{\dr} \sum_{j = 0}^{\dr - i} \sum_{s = 0}^{\db - 1} \sum_{t = 0}^{\db - 1 - s} \bf_1 \lmb (i, j, s, t), (d_r, d_b - 1), \ell \rmb
    \label{eq:qbolone}
    \\
    \qbtlOne &= \sum_{\bd} \frac{\db\pd}{\lab \db \rab} \sum_{i = 0}^{\dr} \sum_{j = 0}^{\dr - i} \sum_{s = 0}^{\db - 1} \sum_{t = 0}^{\db - 1 - s} \bf_2 \lmb (i, j, s, t), (d_r, d_b - 1), \ell \rmb,
    \label{eq:qbtlone}
\end{align}
where for $k=1, 2$, we define
\begin{align}
    \bf_k \lmb (i, j, s, t), (x, y), \ell \rmb &= {x \choose i}{x - i \choose j} \qrol^{i}\qrtl^{j}(1 - \qrol - \qrtl)^{x - i - j} \notag \\ &~~~~ \times {y \choose s}{y - s \choose t} \qbol^{s}\qbtl^{t}(1 - \qbol - \qbtl)^{y - s - t} \times F_k\lmb\lsb i, j, s, t \rsb, (x, y)\rmb.
    \label{eq:f}
\end{align}
\end{widetext}

Equations (\ref{eq:qrolone}) - (\ref{eq:qbtlone})  form a non-linear system,
which can be solved recursively to obtain the steady-state values (i.e., fixed points), $\qrolInf$, $\qrtlInf$, $\qbolInf$, and $\qbtlInf$. Since our goal is to compute the expected size of global cascades {\em given that they exist}, we can initialize this dynamical system with $q_{r,1,0},q_{r,2,0},q_{b,1,0},q_{b,2,0} > 0$.
Because these fixed points account for the probability of being in a corresponding state for the children of the top node, we can use them to calculate the expected size of global cascades.
The expected cascade size stands for the final fraction of active and hyper-active individuals in the network.
This fraction is equal to the probability that the node at the top of the tree turns active or hyper-active.

We give the expected size of the cascades (given that they exist)
in (\ref{eq:sSize}).
The validity of (\ref{eq:sSize}) can be seen as follows:
First, we randomly choose a node, whose colored degree is $\bd = (\dr, \db)$,ƒ with probability $\pd$.
The probability that each of its $\dr$ neighbors (via red links) is active (resp. hyper-active) is given by $\qrolInf$ (resp. $\qrtlInf$).
Similarly, each of the $\db$ neighbors (connected via blue links) of this randomly chosen node is active with probability $\qbolInf$ and hyper-active with probability $\qbtlInf$, independently from each other.
Then, with each possible combination of numbers of active and hyper-active neighbors, we can calculate the probability of being active or hyper-active for the node by the response function (\ref{eq:response_function_multistage2}).
Taking the expectation with respect to the degree $\bd$ yields (\ref{eq:sSize}).
As discussed in details in \cite{gleeson2007seed, yaugan2012analysis, melnik2013multi}, this method, based on the {\em tree-approximation} technique, gives precise results in the asymptotic limit $n \to \infty$, when the underlying network is {\em sparse} and is generated according to the configuration model.
We present extensive numerical studies in Section \ref{sec:numerical_results} that supports our results in the finite node regime.

\begin{widetext}
\begin{align}
  \lim_{n \to \infty}  \mathbb{E}\left[S ~|~ S > 0 \right] = \sum_{\bd}\pd  \sum_{i = 0}^{\dr}\sum_{j = 0}^{d_r - i} \sum_{s = 0}^{\db} \sum_{t = 0}^{\db - s} \left\{ \bf_1 \lmb (i, j, s, t), (\dr, \db), \infty \rmb + \bf_2 \lmb (i, j, s, t), (\dr, \db), \infty \rmb\right\}.
    \label{eq:sSize}
\end{align}
\end{widetext}

From the  recursive equations derived above, we can also obtain the conditions needed for the  global cascades to be possible; i.e., conditions under which $S>0$ with a {\em positive} probability in the limit $n \to \infty$. 
For notational convenience, we define $q_1 \coloneqq \qrolInf$, $q_2 \coloneqq \qrtlInf$, $q_3 \coloneqq \qbolInf$, and $q_4 \coloneqq \qbtlInf$.
Then, the four recursive equations (\ref{eq:qrolone}) - (\ref{eq:qbtlone}) take the form
\begin{align}
    q_i = f_i(q_1, q_2, q_3, q_4), \quad i  = 1, 2, 3, 4.
    \label{eq:recursive}
\end{align}
By direct inspection, we see that the recursive equations (\ref{eq:recursive}) have a trivial fixed point $\qrolInf = \qrtlInf = \qbolInf = \qbtlInf = 0$, which yields $S = 0$ almost surely; this can be seen from the fact that in that case we have $\bbE{S}=0$. In other words, if (\ref{eq:recursive}) has only the trivial fixed point, then global cascades are not possible to take place. 
In general, the trivial fixed point may not be stable and there may exist other non-\textit{trivial} fixed points which can yield $\bbE{S} > 0$; i.e., $S>0$ with a positive probability in the limit $n \to \infty$ and hence global cascade may take place.
To check the existence of non-trivial solutions of (\ref{eq:qrolone}) - (\ref{eq:qbtlone}), we can linearize these equations at $q_1 = q_2 = q_3 = q_4 = 0$ which yields the Jacobian matrix $\mathbf{J}$ given as
\begin{align}
    \mathbf{J} = \frac{\partial f_i(q_1, q_2, q_3, q_4)}{\partial q_j} \Bigg|_{q_1 = q_2 = q_3 = q_4 = 0}.
\end{align}
If the spectral radius, i.e., the largest eigenvalue in absolute value, of the Jacobian matrix is larger than one, then the trivial fixed point $q_1 = q_2 = q_3 = q_4 = 0$ is not stable.
That is, there exist a non-trivial fixed point indicating that global cascades are possible and $S>0$ with positive probability. Otherwise, if the spectral radius of $\mathbf{J}$ is less than or equal to one, then there will be no global cascades.

After the analysis and discussion on the expected size and the conditions of global cascades, we focus next on the probability of triggering a global cascade. In other words, we will calculate the exact (asymptotic) probability that a node selected uniformly at random and turned active leads eventually to a global cascade; i.e., we will compute
\[
\lim_{n \to \infty} \mathbb{P}[S>0].
\]

\subsection{Probability of triggering a global cascade}
\label{sec:prob_trigger_cascade}

We now turn our attention to computing the probability $\mathbb{P}[S>0]$ of global cascades.
As discussed in \cite{watts2002simple, yaugan2012analysis}, the possibility of a seed node to trigger a \textit{global} cascade is closely tied to the size of (and the seed node's connectivity to) the set of \textit{vulnerable} nodes in the network; a node is deemed {\em vulnerable} if it can be activated by only one $\ta$ neighbor \cite{watts2002simple, gleeson2008cascades, dodds2009analysis, yaugan2012analysis}. 
The definition of vulnerable nodes and of the vulnerable component has been extended in \cite{yaugan2012analysis} to the case of multiplex networks. There, a \lq\lq vulnerable component" is defined as a set of nodes, each of which is vulnerable w.r.t. at least one of the link types, such that in the subgraph containing this set of nodes, activating any node leads to the activation of all nodes in the set. A multiplex network is said to contain a {\em giant} vulnerable component (GVC) if the fraction of nodes in its largest vulnerable component is positive in the limit $n \to \infty$. These definitions were then used \cite{yaugan2012analysis} to demonstrate that an initial node can trigger a global cascade if and only if it belongs to 
the {\em extended} giant vulnerable cluster (EGVC), that contains nodes in the GVC and nodes whose activation leads to activation of a node in GVC. Put differently, the probability of a randomly selected node triggering a global cascade is equal to the fractional size of the EGVC; see \cite{yaugan2012analysis} for details.

Here, we use the ideas mentioned above to calculate the probability of global cascades, or equivalently the fraction of nodes that are in EGVC. This will be done through the analysis of a branching process that 
starts from a randomly selected and activated initial node, and keeps exploring the neighboring nodes that are activated according to nodes' response function (\ref{eq:response_function_multistage2}).
The branching process will continue by exploring the neighbors of the newly activated nodes that will also be activated, and so on. The size $S$ of the influence cascade will then be equal to the fraction of nodes identified by this branching process.
\begin{figure}[!t]
	\centering
   \subfigure[An illustration of $G(x)$.]{
        \includegraphics[width=0.22\textwidth]{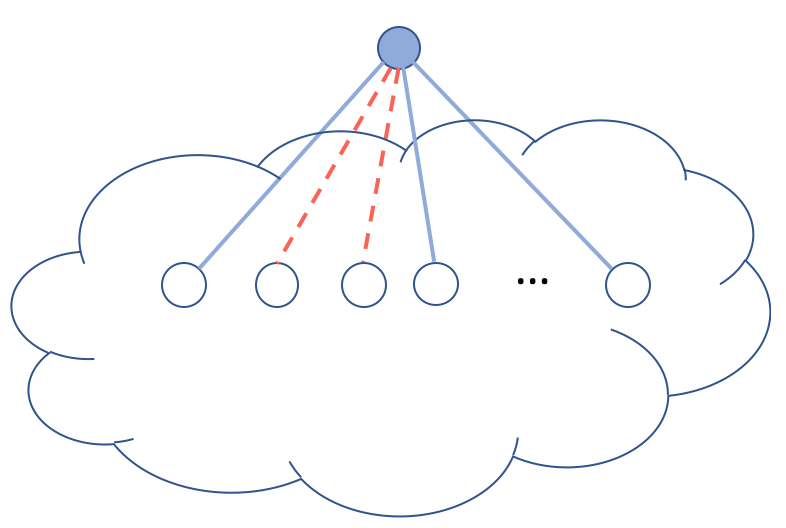}
    }
   \subfigure[An illustration of $\grOne$]{
        \includegraphics[width=0.22\textwidth]{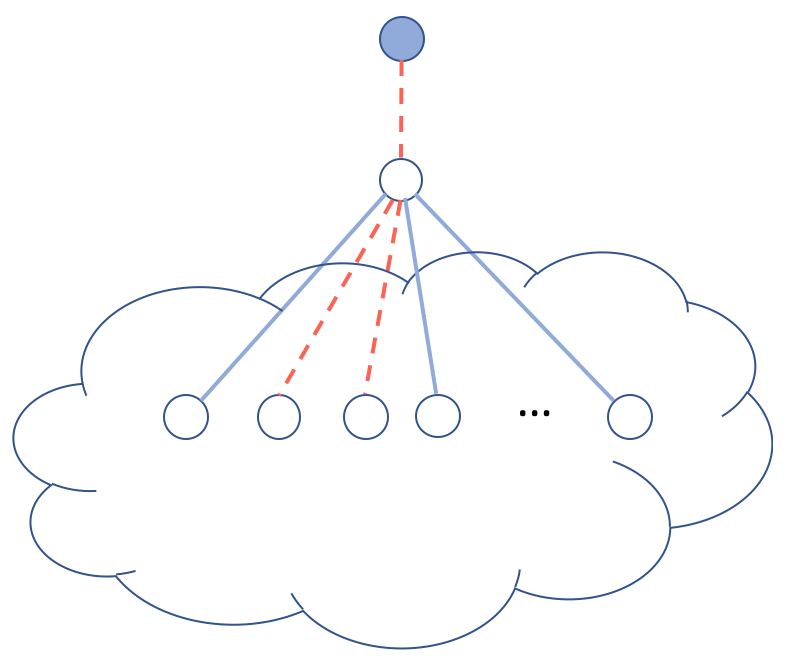}
    }
    \caption{\sl
    The difference between $G(x)$ and $\grOne$.
    Red dashed lines account for red edges in our analysis, while other lines represent blue edges.
    Circles with solid fill indicate active nodes while circles without fill account for  inactive nodes (that can potentially be made active).
    Then, $G(x)$ generates the distribution of the number of active nodes by following the initially activated node, while $\grOne$ generates the distribution of the number of active nodes by following a randomly chosen red edge.
        }
    \label{fig:capitalSmallGx}
\end{figure}

The fraction of nodes identified by the branching process described above can be analyzed using the method of probability generating functions \cite{wilf2013generating}; e.g., see \cite{newman2001random, newman2002spread, watts2002simple, yaugan2012analysis, zhuang2016information} where this tool was demonstrated to be useful for similar purposes. 
The first generating function we use in our analysis is $G(x)$, and it generates the probability distribution of ``the finite number of nodes reached and influenced by the above branching process''; different from \cite{yagan2013conjoining,zhuang2016information,PhysRevE.95.012312}, we exclude the initially activated nodes.
We have
\begin{align}
    G(x) = \sum_{\bd}\pd \grOne^{\dr} \gbOne^{\db},
    \label{eq:capital_gx}
\end{align}
where $\grOne$ (resp. $\gbOne$) generates the probability distribution of ``the finite number of nodes reached and influenced by following a randomly chosen $\tr$ (resp. $\tb$) edge one of whose ends is set to active.''
The difference between $G(x)$ and $\grOne$ is illustrated in Figure \ref{fig:capitalSmallGx}.
The validity of expression (\ref{eq:capital_gx}) can be seen as follows.
First, we initially activate a node which is chosen uniformly at random.
The probability that this node has a degree $\bd = \left(d_r, d_b\right)$ is $\pd$.
In that case, the number of nodes that are reached and activated by this node will be generated (in view of the {\em powers property} of the generating functions) by $\grOne^{\dr} \gbOne^{\db}$. 
Summing over all possible degrees $\bd$ of the initial node leads to (\ref{eq:capital_gx}).

For $(\ref{eq:capital_gx})$ to be useful, we also need to derive expressions for $\grOne$ and $\gbOne$. This will done by the help of two more generating functions. Namely, let $\grTwo$ (resp.~$\gbTwo$) generate the distribution of ``the finite number of nodes reached and influenced by following a $\tr$ (resp.~$\tb$) edge whose one end is connected to a hyper-active node.''
The detailed expressions of the four generating functions are given in (\ref{eq:gr1}) - (\ref{eq:gb2}).
Here, we only explain the derivation $\grOne$, as others can be explained in a similar manner.

To see why (\ref{eq:gr1}) holds, 
first note that as the randomly selected red edge whose one end is connected to an active node is followed, we will find a node with colored {\em degree} $\bd = (d_r, d_b)$
with probability $\frac{\dr\pd}{\lab\dr\rab}$ as already explained in the derivation of  (\ref{eq:qrolone}). There are three possible cases for this node with degree $\bd = (d_r, d_b)$:
\begin{itemize}
    \item It turns \textit{active}, i.e., $\tau_1 \leq \frac{c}{cd_r + d_b} < \tau_2$, which happens with probability $F_1 \lmb (1, 0, 0, 0), (d_r, d_b)\rmb$. Then, this newly activated node will activate $\grOne^{\dr - 1}\gbOne^{\db}$ other nodes based on the powers property of generating functions.
    The reason why we use $\dr - 1$ instead of $\dr$ is because one of its $\dr$ edges has already been considered as its connection to the $\ta$ end.
    \item  It turns \textit{hyper-active}, i.e., $\frac{c}{cd_r + d_b} \geq \tau_2$, which happens with probability $F_2 \lmb (1, 0, 0, 0), (d_r, d_b)\rmb$. Then, the number of nodes reached and influenced by this newly activated node will be generated by $\grTwo^{\dr - 1}\gbTwo^{\db}$; this can be seen via similar arguments to the case above.
    \item It remains inactive, i.e., $\frac{c}{cd_r + d_b}< \tau_1$, which happens with probability $1-F_1 \lmb (1, 0, 0, 0), (d_r, d_b)\rmb-F_2 \lmb (1, 0, 0, 0), (d_r, d_b)\rmb$. Then, there will be no newly activated nodes.
\end{itemize}
Combining these three cases and summing over all possible $\bd$, we get (\ref{eq:gr1}), 
where the explicit factor $x$ accounts for the initial node that is activated. 
The expressions for $\grTwo$, $\gbOne$, and $\gbTwo$ can be derived similarly.

\begin{widetext}
{\small
\begin{align}
g_{r, 1}(x) &= x\sum_{\bd}\frac{\dr\pd}{\lab\dr\rab} \lmb F_1 \lmb (1, 0, 0, 0), (d_r, d_b) \rmb\grOne^{\dr - 1}\gbOne^{\db} + F_2 \lmb (1, 0, 0, 0), (d_r, d_b) \rmb \grTwo^{\dr - 1}\gbTwo^{\db}\rmb \notag\\ & ~~~~ + x^{0}\sum_{\bd}\frac{d_r\pd}{\lab\dr\rab}\left(1 - F_1 \lmb (1, 0, 0, 0), (d_r, d_b) \rmb - F_2 \lmb (1, 0, 0, 0), (d_r, d_b) \rmb \right)
\label{eq:gr1}
\\
g_{r, 2}(x) &= x\sum_{\bd}\frac{\dr\pd}{\lab\dr\rab} \lmb F_1 \lmb (0, 1, 0, 0), (d_r, d_b) \rmb \grOne^{\dr - 1}\gbOne^{\db} + F_1 \lmb (0, 1, 0, 0), (d_r, d_b) \rmb \grTwo^{\dr - 1}\gbTwo^{\db}\rmb \notag\\ & ~~~~ + x^{0}\sum_{\bd}\frac{d_r\pd}{\lab\dr\rab}\left(1 - F_1 \lmb (0, 0, 1, 0), (d_r, d_b) \rmb - F_2 \lmb (0, 0, 1, 0), (d_r, d_b) \rmb \right)
\label{eq:gr2}
\\
g_{b, 1}(x) &= x\sum_{\bd}\frac{\db\pd}{\lab\db\rab} \lmb  F_1 \lmb (0, 0, 1, 0), (d_r, d_b) \rmb \grOne^{\dr}    \gbOne^{\db - 1} + F_2 \lmb (0, 0, 1, 0), (d_r, d_b) \rmb \grTwo^{\dr}\gbTwo^{\db - 1}\rmb \notag\\ & ~~~~ + x^{0}\sum_{\bd}\frac{d_b\pd}{\lab\db\rab}\left(1 - F_1 \lmb (0, 0, 1, 0), (d_r, d_b) \rmb F_2 - \lmb (0, 0, 1, 0), (d_r, d_b) \rmb \right)
\label{eq:gb1}
\\
g_{b, 2}(x) &= x\sum_{\bd}\frac{\db\pd}{\lab\db\rab} \lmb F_1 \lmb (0, 0, 0, 1), (d_r, d_b) \rmb  \grOne^{\dr}    \gbOne^{\db - 1} +  F_2 \lmb (0, 0, 0, 1), (d_r, d_b) \rmb \grTwo^{\dr}\gbTwo^{\db - 1}\rmb \notag\\ & ~~~~ + x^{0}\sum_{\bd}\frac{d_b\pd}{\lab\db\rab}\left(1 - F_1 \lmb (0, 0, 0, 1), (d_r, d_b) \rmb - F_2 \lmb (0, 0, 0, 1), (d_r, d_b) \rmb\right)
\label{eq:gb2}
\end{align}
}
\end{widetext}

These recursive equations can be used to compute the probability that a global cascade is triggered in the following manner. Since $G(x)$ generates the number of {\em finite} nodes reached and activated by this branching process, we should have $G(1)=1$ by the conservation of probability, {\em unless} there is a positive probability that the branching process leads to an {\em infinite} number of nodes. In other words, $1-G(1)$ corresponds to the probability that the branching process under consideration will {\em survive} forever and will not go extinct, meaning that the underlying influence propagation process will constitute a global cascade. Thus, we have
\begin{align}
\lim_{n \to \infty} \mathbb{P}[S>0] = 1-G(1).
\label{eq:prob_vs_G_1}
\end{align}
This approach has been introduced in \cite{watts2002simple} and used in \cite{newman2001random,yaugan2012analysis} for similar calculations.

In order to calculate $G(1)$, we now solve for the fixed point of (\ref{eq:gr1})-(\ref{eq:gb2}) at  $x = 1$. Simplifying the notation as $g_1 \coloneqq g_{r, 1}(1)$, $g_2 \coloneqq g_{r, 2}(1)$, $g_3 \coloneqq g_{b, 1}(1)$, and $g_4 \coloneqq g_{b, 2}(1)$,
the recursive equations  (\ref{eq:gr1})-(\ref{eq:gb2}) at $x = 1$ can be expressed as 
\begin{align}
    g_i = h_i(g_1, g_2, g_3, g_4), \qquad i = 1, 2, 3, 4.
    \label{eq:simple_recursive_equation}
\end{align}
Here, the exact form of the functions $h_1(g_1, g_2, g_3, g_4), \ldots, h_4(g_1, g_2, g_3, g_4)$ will be obtained from (\ref{eq:gr1})-(\ref{eq:gb2}). Once the fixed points of (\ref{eq:simple_recursive_equation}) are obtained, we  
get from (\ref{eq:capital_gx}) that 
\begin{align}
    G(1) = \sum_{\bd}\pd g_1^{\dr} g_3^{\db}.
    \label{eq:capital_g1}
\end{align}
In view of (\ref{eq:prob_vs_G_1}), we finally obtain the desired probability of global cascades as 
\begin{align}
\lim_{n \to \infty} \mathbb{P}[S>0] = 1-\sum_{\bd}\pd g_1^{\dr} g_3^{\db}.
\label{eq:prob_trigger}
\end{align}


\section{Numerical Results}
\label{sec:numerical_results}
In this section, we present numerical results to support our analysis on the expected  size and  probability of global cascades. In particular, we are interested in checking the accuracy of our asymptotic results when the number of nodes is finite.
We will also investigate via extensive simulations the impact of hyper-influencers (i.e., the additional influence exerted by them) on the contagion dynamics.

\subsection{The agreement between our analysis and simulations}
\label{sec:agreement_ana_exp}
\begin{figure}[t]
	\centering
   \subfigure[]{
        \includegraphics[width=0.5\textwidth]{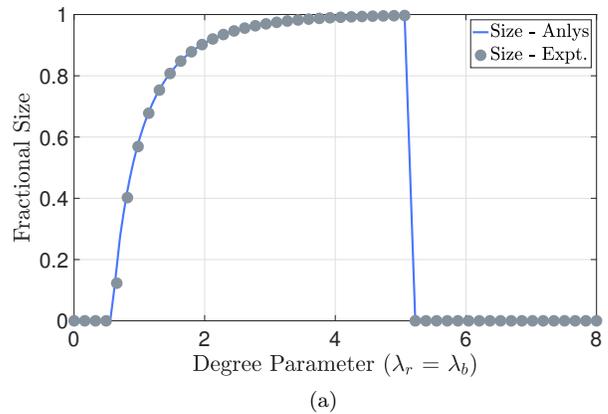}
    }
    \caption{\sl
        Simulations for doubly Poisson degree distributions,  $n = 1 \times 10^{6}$, $\alpha = 0.5$, $\tau_1 = 0.18$, and $\tau_2 = 0.32$.
        The weight of hyper-influencers is taken to be $\beta = 1.5$.
        }
    \label{fig:ana_exp_agreement_size}
\end{figure}

First, we focus on demonstrating the accuracy of our analytic results on the expected size of global cascades and the probability of having global cascades in the finite node regime.
In our numerical simulations, we use a doubly Poisson distribution to assign the number of red and blue edges for each node.
Namely, with $p_{k}^{r}$ (resp. $p_{k}^{b}$) denoting the probability that a node is assigned $k$ red (resp. blue) edges, we let 
\begin{align}
p_{k}^{b} &= e^{-\lambda_{b}}\frac{(\lambda_{b})^{k}}{k!}, \mbox{ \ k = 0, 1, \dots },
\label{eq:poisson1}
\\
p_{k}^{r} &= \alpha e^{-\lambda_{r}}\frac{(\lambda_{r})^{k}}{k!} +  (1 - \alpha) \delta_{k, 0}, \mbox{ \ k = 0, 1, \dots }.
\label{eq:poisson2}
\end{align}
Here, $\lambda_r$ (resp. $\lambda_b$) denotes the mean number of red (resp. blue) edges assigned per node, $\alpha$ denotes the fraction of nodes that have red edges (i.e., the relative size of the {\em red} network $\mathbb{R}$), and $\delta$ denotes the Kronecker delta.
In our simulations to verify our analysis on the expected size, we use $n = 1 \times 10^{6}$ nodes\footnote{To avoid the finite size effect, we use $n = 2 \times 10^{6}$ as the number of nodes around the second phase transition in the simulations.} to create networks and set $\alpha = 0.5$.
Besides, we use $c = 0.5$ and $\beta = 1.5$ as the content parameter and the weight of hyper-active nodes, respectively, and fix $\tau_1 = 0.18$ and $\tau_2 = 0.32$.
Then, for several values of $\lambda_r = \lambda_b$, we run 1,000 independent experiments (for each parameter set), each time computing the fraction of nodes that eventually turn active or hyper-active.
The results are depicted in Figure \ref{fig:ana_exp_agreement_size} where lines represent analytical results obtained from (\ref{eq:qrolone}) - (\ref{eq:qbtlone}), and symbols represent the average cascade size obtained in simulations (over 1,000 experiments for each data point).
We see that there is a good agreement between the analytic results and the simulations. 

\begin{figure}[t]
	\centering
   \subfigure[]{
        \includegraphics[width=0.5\textwidth]{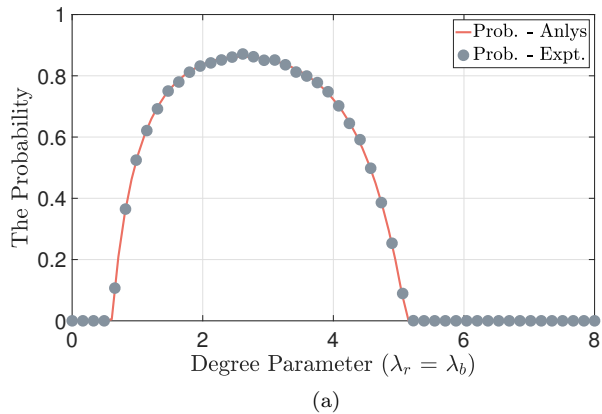}s
    }
    \caption{\sl
        Simulations for doubly Poisson degree distributions,  $n = 1 \times 10^{6}$, $\alpha = 0.5$, $\tau_1 = 0.18$, and $\tau_2 = 0.32$.
        The weight of hyper-influencers is taken to be $\beta = 1.5$.
        }
    \label{fig:ana_exp_agreement_prob}
\end{figure}

Next, to check the correctness of our analysis on the probability, we fix all parameters except increasing the number of experiments from 1,000 to 10,000.
As shown in Figure \ref{fig:ana_exp_agreement_prob}, we observe that our analysis on the probability (\ref{eq:gr1}) - (\ref{eq:gb2}) also match very well the simulations results.
This indicates that, although asymptotic in nature, the results presented in Section \ref{sec:results} are still helpful in understanding complex contagion dynamics (e.g., the probability and expected size of global cascades) in finite networks. 

In addition,  we  observe from both Figure \ref{fig:ana_exp_agreement_size} and \ref{fig:ana_exp_agreement_prob}  that the contagion exhibits two {\em phase transitions}, i.e., two different $\lambda_r = \lambda_b$ values around which fractional cascade size transitions from zero to a positive value, or vice verse.
These points are of great interest since they provide insights on how network connectivity affects the possibility of observing global influence spreading events.
The first transition occurs around {\em low} values of $\lambda$, and reflects the fact that global spreading events become possible only after the network reaches a certain level of connectivity.
The second phase transition occurs around {\em high} $\lambda$ values, indicating that global cascades can not occur when nodes are locally stable; i.e., when they have a large number of friends, individuals tend to be difficult to get influenced by a few active neighbors.

After demonstrating the correctness of our analysis, we  focus on exploring the impact of hyper-influencers on complex contagion dynamics in the following sections.

\subsection{The impact of hyper-influencers on the global cascade boundary}
In this section, we investigate
how the parameters $\beta, \tau_1, \tau_2$ of the contagion model and the connectivity of the network jointly affect the possibility of global cascades. In particular, we will determine the boundaries in the space of parameters that separate the region where global cascades are possible (i.e., $\mathbb{P}[S>0]$) from the region where global cascades do {\em not} take place almost surely (i.e., $\mathbb{P}[S=0]$). First, we will focus on the impact of the weight $\beta$ of hyper-influencers  on the global cascade boundary,
and then move on to the discussion about the impact of the threshold $\tau_1$ of ordinary influencers.

\begin{figure}[t]
	\centering
    \includegraphics[width=0.5\textwidth]{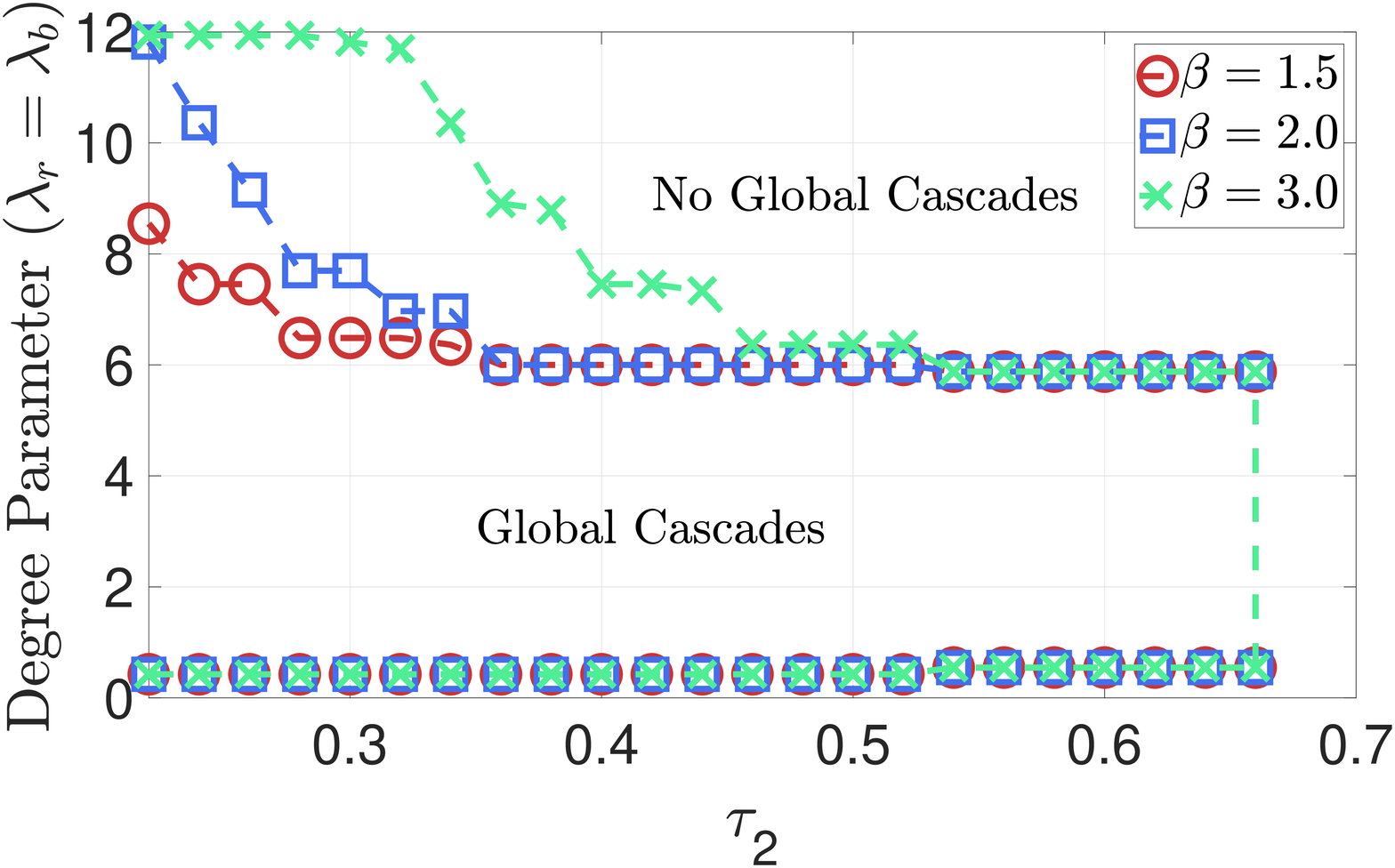}
    \caption{\sl
        Given $\tau_1 = 0.15$ and $\alpha = 0.5$, we vary the mean degree $\lambda$ and $\tau_2$ to plot the global cascade region for several $\beta$, 1.5, 2.0, and 3.0.
        }
	\label{fig:fix_tau1_vary_beta}
\end{figure}

Figure \ref{fig:fix_tau1_vary_beta} shows the global cascade boundary in the space of 
 $\tau_2$ and degree parameter $\lambda=\lambda_r=\lambda_b$, for 
 several values of $\beta$. 
We observe that larger $\beta$ values lead to a larger region of parameters $\tau_2,\lambda$ for which global cascades can take place; i.e., the global cascade region gets larger with increasing $\beta$.
An interesting observation is that the cascade boundary is more sensitive to the changes in $\beta$ values when $\lambda$ is {\em large}; i.e., the lower parts of the boundaries seen in Figure \ref{fig:fix_tau1_vary_beta} are less dependent on the choice of $\beta$ as compared to the upper parts.  
This can be explained as follows.
When $\lambda$ is {\em small}, the existence of global cascades (and hence the cascade boundary) is mainly determined by whether the network has enough connectivity to spread the influence. However, increasing $\beta$ does not change the connectivity of the network, and hence does not affect the boundary when $\lambda$ is low.
Differently, when $\lambda$ {\em high}, the location of the boundary (i.e., the second phase transition points seen in Figures \ref{fig:ana_exp_agreement_size}-\ref{fig:ana_exp_agreement_prob}) is decided by the likelihood of nodes with high degree being influenced by a single active or hyper-active neighbor. Thus, the boundary is determined from a node's perceived influence, or perceived proportion of active and hyper-active neighbors, given at (\ref{eq:response_function_multistage2}), and on how this compares with the activation thresholds $\tau_1$ and $\tau_2$. From (\ref{eq:response_function_multistage2}), we see that higher $\beta$ leads to an increased perceived influence for a node that has at least one hyper-active neighbor, making it possible for the activation threshold to be exceeded at higher  $d_r,d_b$ values (equivalently at higher $\lambda$ values). 
Thus, when $\lambda$ is {\em high}, the boundary tends to be more sensitive to the changes in $\beta$.

\begin{figure}[t]
	\centering
    \includegraphics[width=0.5\textwidth]{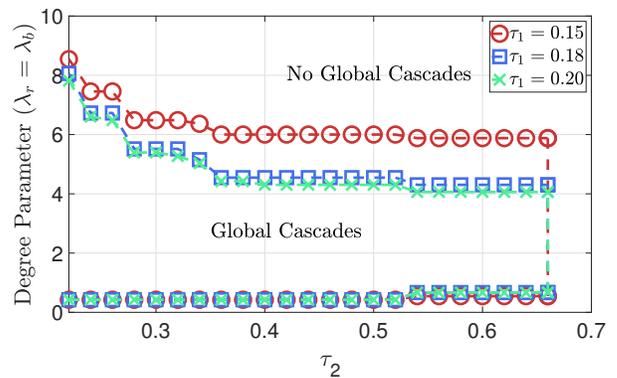}
    \caption{\sl
        Given the weight of extra influence $\beta = 1.5$ and $\alpha = 0.5$, we vary the  degree parameter $\lambda=\lambda_b=\lambda_r$ and $\tau_2$ to plot the region where there exist a global cascade for severl $\tau_1$, 0.15, 0.18, and 0.2.
        Both of the edges are assigned by the doubly Poisson distribution in Section \ref{sec:agreement_ana_exp}.
        }
	\label{fig:fix_beta_vary_tau1}
\end{figure}

Next, we investigate the impact of the activation threshold $\tau_1$ on the global cascade boundary (again considering the space of $\lambda-\tau_2$.
In Figure \ref{fig:fix_beta_vary_tau1}, we fix $\beta = 1.5$ and plot the boundary on the $\tau_2-\lambda$ plane that separates the regions where cascades are possible and not possible, respectively. This is done for three different values of $\tau_1$.
We observe that the impact of $\tau_1$ (i.e., the threshold on the perceived influence that an inactive node needs to receive in order to turn active) on the cascade boundary is opposite to that of $\beta$.
That is, the higher $\tau_1$ is, the smaller is the region where \textit{global} cascades are possible.
The reason behind this observation is as follows.
From the expression of the response function (\ref{eq:response_function_multistage2}), we see that it is decreasing with increasing $\tau_1$. 
In other words, a higher $\tau_1$ makes it harder for nodes to become active (i.e., influenced), leading to a smaller cascade region.

\subsection{The impact of hyper-influencers on the probability and expected size of global cascades}
\label{sec:impact_extra_influence_hyper}

\begin{figure}[t]
	\centering
    \includegraphics[width=0.5\textwidth]{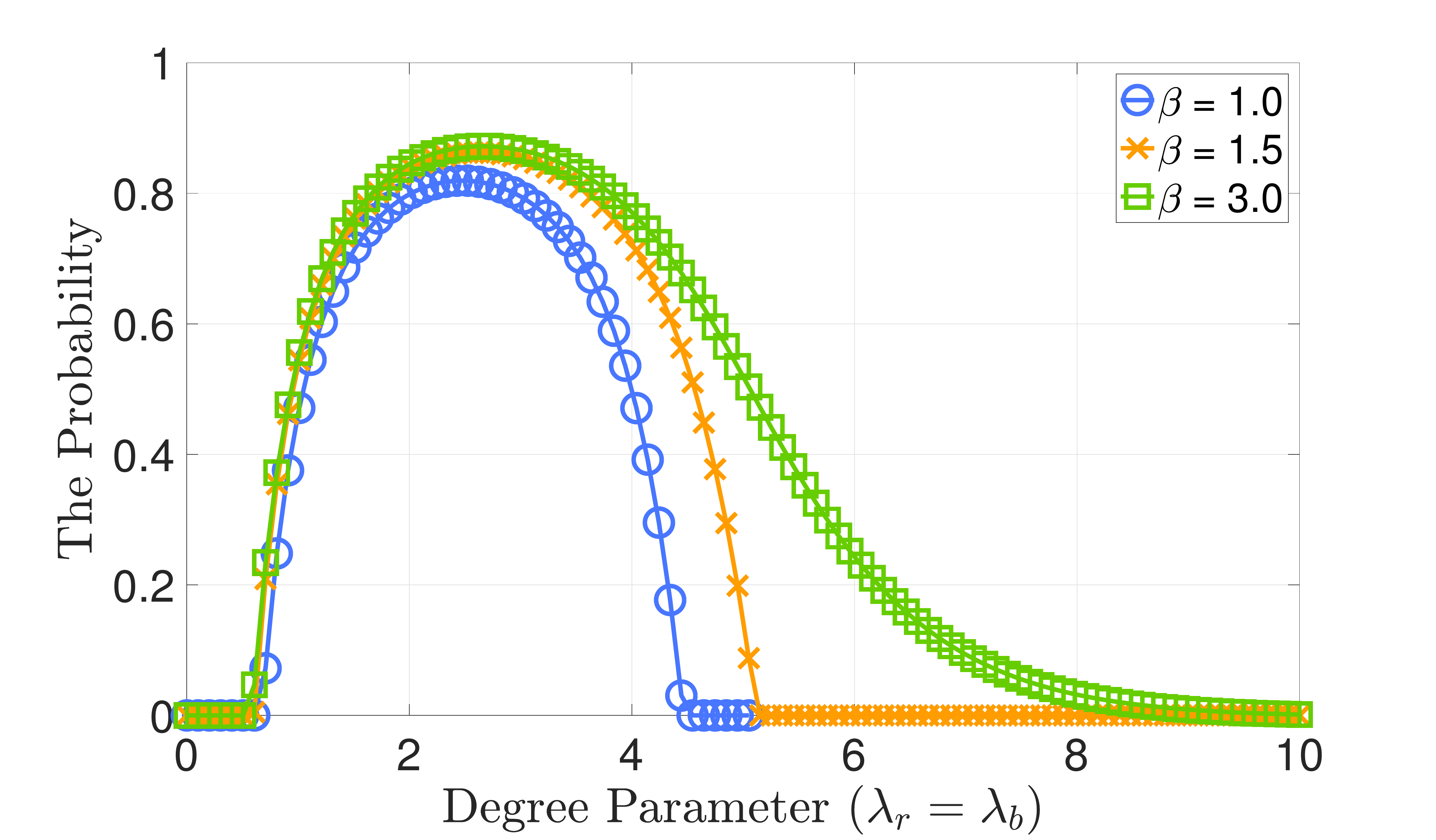}
    \caption{
        The comparison between different $\beta$ for the probability of triggering a $\tg$ cascade.
        }
	\label{fig:comparison_between_beta_prob}
\end{figure}

We start by investigating the impact of the extra influence $\beta$ (that hyper-active nodes exert on their neighbors) on the probability of global cascades. From Figure \ref{fig:comparison_between_beta_prob}, we observe that a larger $\beta$ will increase the probability of triggering a global cascade.
This observation is intuitive given that the response function (\ref{eq:response_function_multistage2}) is increasing with respect to $\beta$.
Thus, with a higher $\beta$, the perceived influence from a single active or hyper-active neighbor exceeds the threshold more easily, leading to a larger \textit{vulnerable} component.
Also, we see in Figure \ref{fig:comparison_between_beta_prob} that when the degree parameter is {\em large}, the cascade probability becomes more sensitive to the changes in $\beta$. This is consistent with the observations from Figure \ref{fig:fix_beta_vary_tau1} and can be explained in a similar manner.

\begin{figure}[t]
	\centering
    \includegraphics[width=0.5\textwidth]{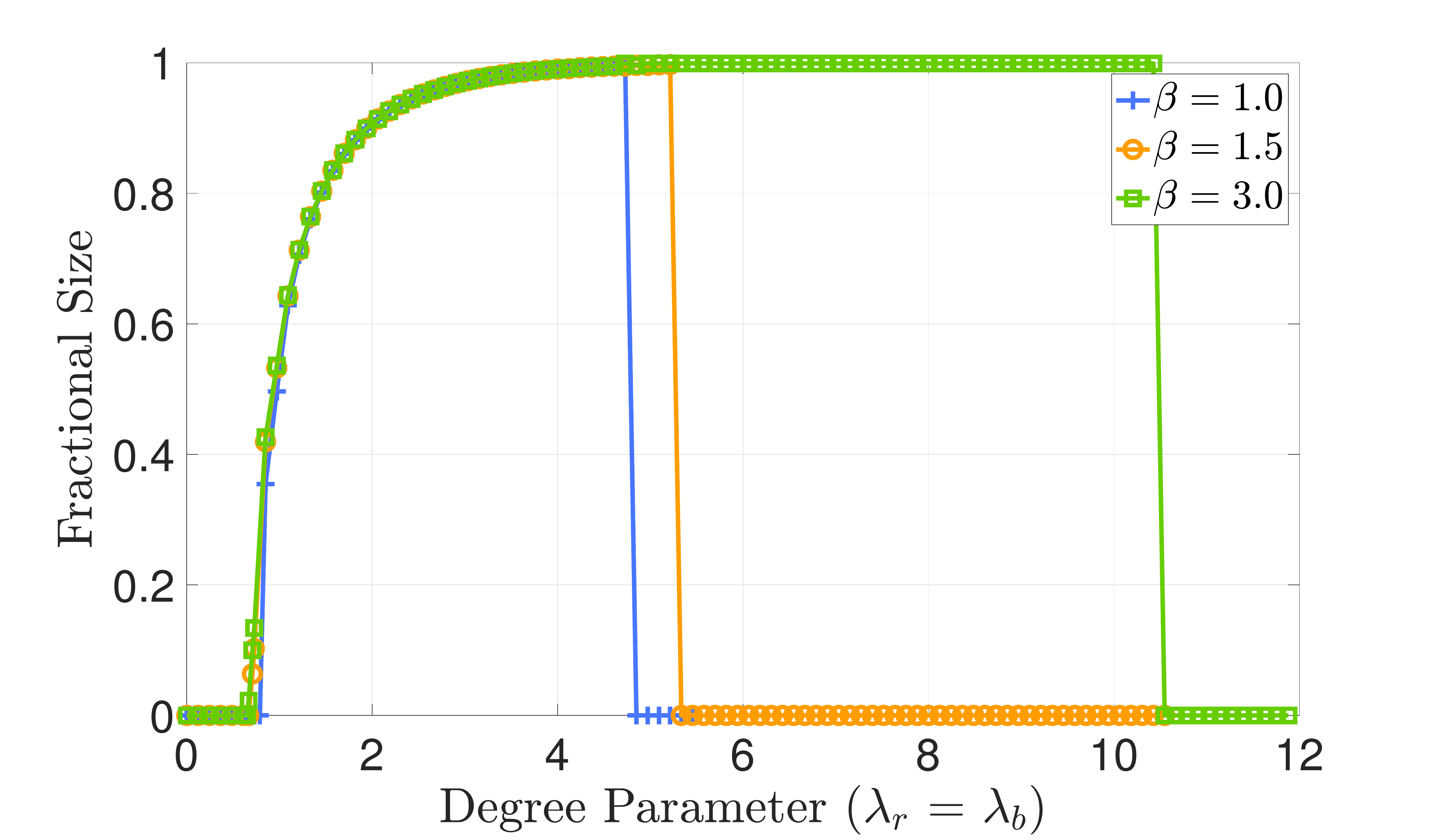}
    \caption{
        The comparison between different $\beta$ for the expected $\tg$ cascade size.
        }
	\label{fig:comparison_between_beta_size}
\end{figure}

Next, we discuss the impact of hyper-influencers on the expected size of global cascades.
From Figure \ref{fig:comparison_between_beta_size}, we observe that increasing $\beta$ leads to an expansion of the interval of $\lambda_r=\lambda_b$ values for which expected  cascade size is positive. However, over the common interval where cascade size is positive, we see that  increasing $\beta$ nearly does not lead to changes in the expected cascade size.
The reason behind this observation is that the expected cascade size is mainly determined by the connectivity, (e.g., the mean degree) of the network, which remains invariant to changes in $\beta$. Thus, increasing $\beta$ nearly does not change the expected size of global cascades.
The expansion of the interval over which $S>0$ with increasing $\beta$ is explained by  the response function (\ref{eq:response_function_multistage2}) being increasing in $\beta$. In other words, a higher $\beta$ makes it easier for the perceived influence to exceed the activation threshold, helping global cascades take place even at higher mean degree.

\subsection{The impact of hyper-influencers in multiplex networks}
In this section, we investigate more closely how hyper-influencers affect the complex contagions.
To this end, we restrict hyper-active nodes to have additional influence only through one type of edges, red or blue, rather than allowing them to exert additional influence through both types of edges.
This setting is motivated by cases where people can reach a more active/influential state only in one network, or one relationship type.
For example, some people may be reluctant to express their opinions freely in person (e.g., physical networks), but may be much more active on online networks (e.g., Twitter) due to anonymity.
This raises an interesting question: which network or edge type would facilitate the influence propagation process most when hyper-influencers are allowed there.
In what follows, we conduct several experiments to answer this question: 1) we only allow hyper-activity in red edges, i.e., hyper-active neighbors connected by blue edges will be counted as merely active when checking the response function; 2) we only allow hyper-activity in blue edges.
More importantly, we conduct these experiments on a network with {\em low} assortativity and then a network with {\em high} assortativity to see if assortativity has any impact on the answer to the above question.
Assortativity is defined as the Pearson correlation coefficient between the degree of nodes that are connected by a link \cite{newman2002assortative}.
If a network is assortative, then nodes of high degree in the network tend to attach to high degree nodes; it was noted in \cite{newman2002assortative} that  social networks tend to have high assortativity.

In the following experiments, we use the degree distributions (\ref{eq:poisson1}) and (\ref{eq:poisson2}) to assign red and blue degrees.
However, to be able to control the assortativity of networks, we set $\alpha\lambda_r = \lambda_b$ rather than $\lambda_r = \lambda_b$.
With this setting, when $\alpha$ is large, e.g., 0.99, nearly all of the nodes will have a similar number of red and blue edges, which leads to networks with limited assortativity.
On the contrary, when $\alpha$ is {\em low}, e.g., 0.1, only 10\% of the nodes will have extra red edges.
In addition, these nodes will have a significantly larger number of edges, since $\lambda_r$ is ten times larger than $\lambda_b$.
The nodes with extra red edges will tend to be connected together, which results in the network to have {\em high} assortativity.
A more detailed discussion on this can be found in \cite{PhysRevE.95.012312}.

\begin{figure}[t]
	\centering
    \includegraphics[width=0.45\textwidth]{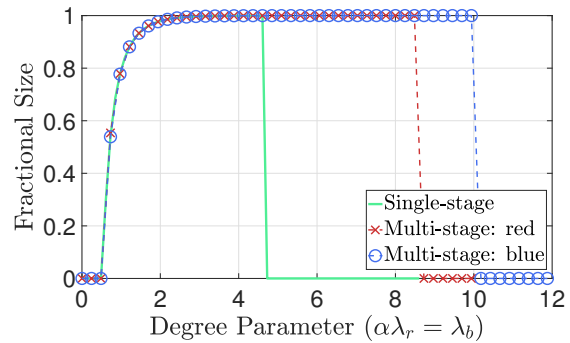}
    \caption{
        Hyper-activity only appears in either red or blue edges.
        We fix $\tau_1 = 0.18$ and $\tau_2 = 0.32$, and vary the mean degree.
        When $\alpha = 0.99$, the assortativity is negligible.
       \vspace{-1mm} }
	\label{fig:only_red_0_vs_10_noass}
\end{figure}

We start with the limited assortativity case, i.e., $\alpha = 0.99$.
As shown in Figure \ref{fig:only_red_0_vs_10_noass}, we observe that regardless of which network hyper-influencers are constrained to exist, there are two phase transitions as in the case of single-stage complex contagions.
However, we see that the existence of hyper-influencers delays the second phase transitions to higher mean degrees.
The reason behind this delay can be explained as follows.
As mentioned before, the second phase transition occurs due to {\em high} local stability of nodes making their states hard to change by only few active neighbors.
However, hyper-influencers help increase the value of the perceived influence, i.e., $\frac{c(m_{r, 1} + \beta m_{r, 2}) + m_{b, 1} + \beta m_{b, 2}}{c d_{r} + d_{b}}$, so that the response function could be exceeded even with few active and hyper-active neighbors, in the high mean degree region.
Besides, allowing hyper-activity in blue edges leads to a larger region where global cascades take place, in comparison with the case where hyper-activity exists only in red edges. 
This can be explained as follows.
When $\alpha = 0.99$, there are more nodes connected by blue edges in the network than red edges.
That is, the impact of blue edges on impeding global cascades is more than that of red edges.
Thus, allowing hyper-influence to be exerted in blue edges delays the second phase transition further.

\begin{figure}[t]
	\centering
    \includegraphics[width=0.45\textwidth]{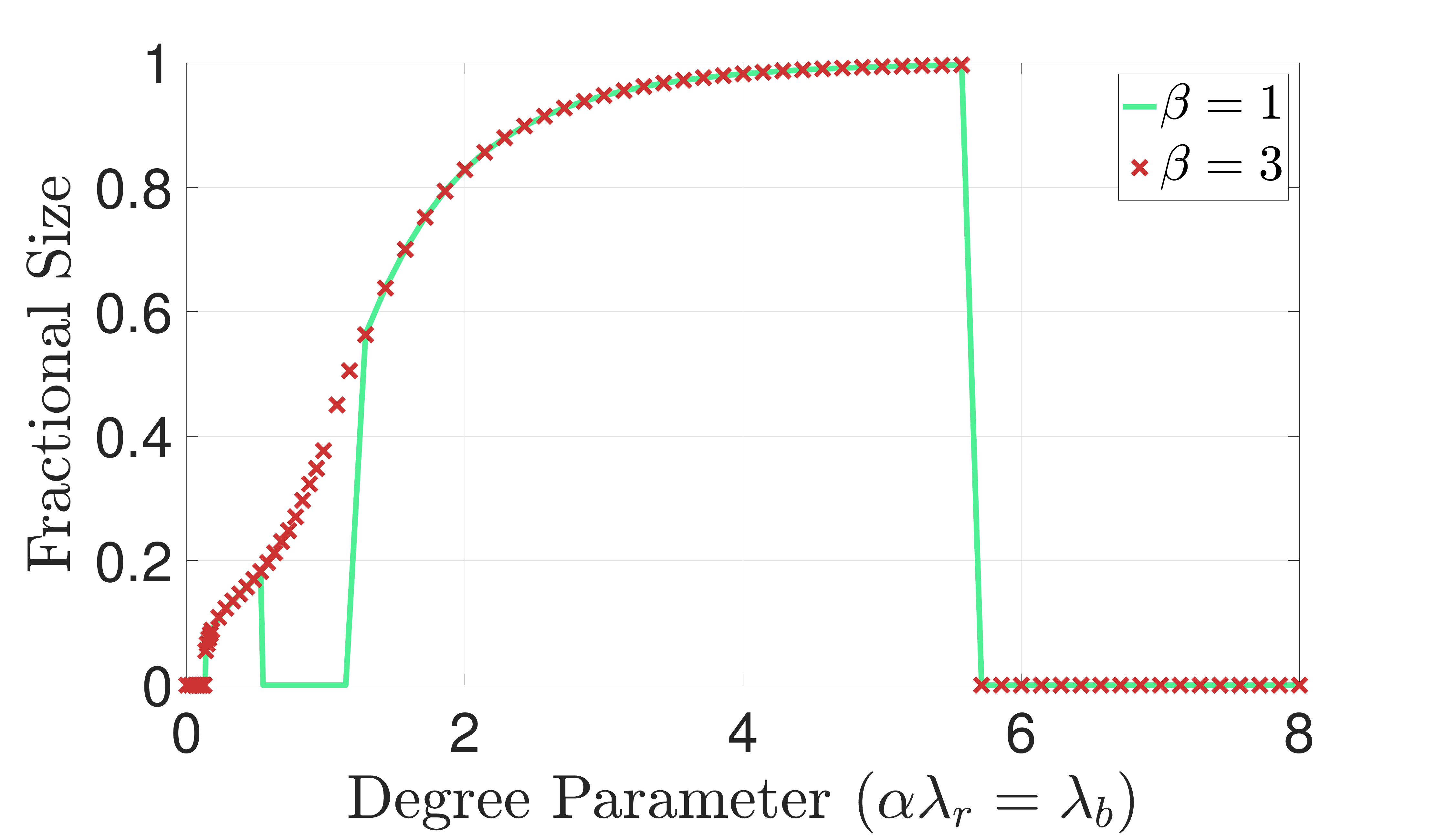}
    \caption{\sl
        Hyper-activity only appears in red edges.
        We fix $\tau_1 = 0.18$ and $\tau_2 = 0.32$, and vary the mean degree.
        When $\alpha = 0.1$, the assortativity of the network is around 0.8.
        }
	\label{fig:only_red_0_vs_10}
\end{figure}

Next, we discuss the case where $\alpha = 0.1$ that leads to a highly assortative network \cite{PhysRevE.95.012312}.
In Figure \ref{fig:only_red_0_vs_10}, we present numerical results for the first setting where the hyper-active state is manifested in only red edges.
When $\beta = 1$, i.e., when there are no hyper-influencers in the network, four phase transitions take place.
However, if we increase $\beta$ from one to three, then we only observe two phase transitions.
This can be explained as follows.
When $\beta = 1$, multi-stage complex contagions is reduced to single-stage complex contagions, in which case four phase transitions might occur when assortativity is high \cite{PhysRevE.95.012312}.
As explained in \cite{PhysRevE.95.012312}, the first pair of phase transitions are mainly due to the red edges.
When $\lambda_b$ is {\em small}, there are too few blue edges to trigger a global cascade.
However, since we have $\lambda_r = 10 \lambda_b$, there are still enough red edges to have global cascades.
As we increase $\lambda_b$, we observe a parameter interval where red edges are too many while blue edges are too few to have a global cascade.
If we keep increasing $\lambda_b$ further, global cascades start appearing again when the network has enough connectivity in blue edges to propagate the influence.
However, further increasing in $\lambda_b$ leads to high local stability of nodes w.r.t. both blue and red edges and global cascades become impossible again.
A more detailed discussion can be found in \cite{PhysRevE.95.012312}.

The reason why increasing $\beta$ changes the number of phase transitions is as follows.
From the definition (\ref{eq:response_function_multistage2}) of the response function, we observe that it is monotonically increasing with respect to $\beta$.
Thus, when $\beta$ is higher, an inactive node is easier to be activated by a hyper-active node, which makes it possible to have global cascades at higher levels of connectivity; i.e., the second phase transition tends to appear at larger $\lambda$.
This leads to the second and the third phase transitions seen in Figure \ref{fig:only_red_0_vs_10} when $\beta=0$ disappear; i.e., the interval where we have too many red and too few blue edges disappears.

\begin{figure}[t]
	\centering
    \includegraphics[width=0.45\textwidth]{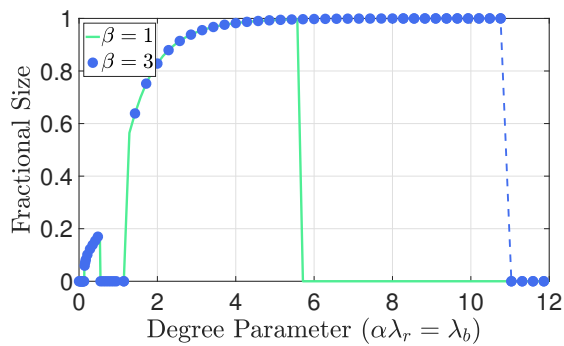}
    \caption{\sl
        Hyper-activity only appears in blue edges.
        We fix $\tau_1 = 0.18$ and $\tau_2 = 0.32$, then vary the mean degree.
        When $\alpha = 0.1$, the assortativity is high (be up to 0.8).
        }
	\label{fig:only_blue_0_vs_10}
\end{figure}

Next, we focus on the second setting where hyper-activity is only manifested in blue edges.
The results are shown in Figure \ref{fig:only_blue_0_vs_10}.
Allowing hyper-activity in blue edges does not change the connectivity of the network, so the first and the second phase transitions caused by the connectivity w.r.t. red edges remain the same.
However, the gap between the second and third transitions still exists.
The gap happens between the second transition w.r.t. red and the first transition w.r.t. blue edges.
A high $\beta$ only shifts the second transition to the right but does not affect the first transition much. 
Thus, the gap disappears quickly with increasing $\beta$ when we allow it in red edges, but remains when we only allow it in blue edges.
Besides, compared with the case $\beta = 1$, the fourth transition is significantly delayed when $\beta = 3$.
The reason behind the delay of the fourth phase transition is similar to the previous discussion:
A higher $\beta$ makes it easier to exceed the threshold even when the degree parameter is at a high level, so the original fourth phase transition has been extended to a larger mean degree.

From these experiments, we  conclude that depending on the assortativity of the network, the impact of hyper-activity in red or blue edges on complex contagions are different: when the network is highly assortative, the additional influence exerted by the hyper-active nodes may change not only the critical transition points, but also the number and order of phase transitions, while for networks that have little or no assortativity, the additional influence mainly enlarge global cascade regions.

\section{Conclusion and Future Work}
\label{sec:conclusion}
In this work, we study the propagation of influence in multiplex networks 
under a {\em multi-stage} complex contagion model.
We derive recursive relations characterizing the dynamics of influence propagation, and compute the probability of triggering \textit{global} cascades and the expected size of \textit{global} cascades, i.e., cases where a single individual can initiate a propagation that eventually influences a positive fraction of the population. The analytic results are also confirmed and supported by an numerical study.
In particular, we demonstrate how the additional influence exerted by the hyper-active nodes can enlarge the network parameter region where global cascades take place. 
An interesting finding is that depending on the assortativity of the network, the existence of hyper-influencers affect the expected size of global cascades differently.
For instance, when the network is highly assortative, the additional influence exerted by the hyper-active nodes may change not only the critical transition points, but also the number and order of phase transitions; while the affect is much more limited in networks with low 
assortativity. 

There are many interesting directions to pursue for future work. 
First, it might be interesting to extend this work to more general network models than the configuration model used here. For instance, it would be interesting to consider networks that have high clustering. It would also be interesting to study multi-stage complex contagions  using non-linear threshold models.
Finally, it would be interesting to consider the case where it is possible for a node to transition back to the inactive state after being activated, e.g., due to the {\em negative} influence received by several {\em hyper-inactive} neighbors. %




\bibliography{yong}
\bibliographystyle{IEEEtran}
\end{document}